\documentstyle[seceq,epsf,preprint]{ptptex}

\newcommand{\la}{\langle}
\newcommand{\ra}{\rangle}

\newcommand{\mr}{{\mib r}}
\newcommand{\mx}{{\mib x}}
\newcommand{\my}{{\mib y}}

\newcommand{\hphi}{{\hat \phi}}
\newcommand{\hpi}{{\hat \pi}}
\newcommand{\hpsi}{{\hat \psi}}
\newcommand{\hQ}{{\hat Q}}
\newcommand{\hP}{{\hat P}}

\notypesetlogo  

\title{
Variational Approach to the Dynamics of \\
Bose-Einstein Condensates. I
}
\subtitle{Variational Ansatz and Quantum Dynamics}

\author{
Y. Tsue$^1$\footnote{E-mail: tsue@cc.kochi-u.ac.jp}, 
D. Vautherin$^2$\footnote{deceased} and 
T. Matsui$^3$\footnote{E-mail: tmatsui@nt1.c.u-tokyo.ac.jp}
}
\inst{
$^1$Physics Division, Faculty of Science, Kochi
University, Kochi 780-8520, Japan \\
$^2$L.P.N.H.E., L.P.T.P.E., Universit\'es Paris VI/VII,
Paris Cedex 05, France \\
$^3$Institute of Physics, University of Tokyo,
Komaba, 
Tokyo 153-8902, Japan
}

\recdate{
\today
}

\abst{
Variational method is applied to describe Bose-Einstein condensates (BEC)
interacting via a pseudo-potential, taking into account quantum
fluctuations around the mean field by the Gaussian approximation.
Contributions from the pair-wise scattering by the singular potential
should be removed carefully to avoid double countings in the
perturbation series in terms of original bare interaction.
We show that this procedure removes all dangerous terms which contain
divergences. 
}

\begin{document}

\maketitle

\section{Introduction}

Recent experimental breakthrough\cite{BEC} of manufacturing Bose-Einstein
condensates (BEC)
in magnetically trapped dilute gases of alkali-metal atoms has created new
opportunities to test many-body theories in well-controlled experimental
conditions.
Although the phenomena can be considered as a spectacular realization of
an old prediction
of quantum statistical mechanics as applied to ideal bose gases\cite{old},
it is known
that mutual interactions of particles still play important roles to determine
characteristic properties of the systems.\cite{BP96,RMP99}

In theoretical treatments the condensate is usually 
described by the dilute gas approximation in which the pair-wise short range
interaction is fully accommodated by the use of the effective interaction,
usually referred to as the pseudo-potential, 
defined by
\begin{equation}\label{1.1}
V_{\rm eff.} (\mr) = \frac{4 \pi \hbar^2 a}{m} \delta ( \mr )
= g \delta ( \mr ) \ ,
\end{equation}
where $a$ is the $S$-wave scattering length of the pair interaction and
$m$ is the mass of the particle,
while the many-body collective dynamics 
of the condensate is described by the Gross-Pitaevskii equation\cite{GP61},
also known as the non-linear Schr\"{o}dinger equation,
for a classical field $\Phi (\mr, t)$;
\begin{equation}\label{GPeq}
i \hbar \frac{\partial \Phi (\mr, t)}{\partial t} = 
\left[ - \frac{\hbar^2 \nabla^2}{2m}  + g | \Phi (\mr, t) |^2 \right]
\Phi (\mr, t) \ .
\end{equation}
This equation can be derived from the Heisenberg equation of motion of the
quantum field ${\psi} (\mr, t)$
\begin{equation}
i \hbar \frac{\partial {\psi} (\mr, t)}{\partial t}
 =  [ {\psi} (\mr, t) , {\hat H} ] \nonumber
 = - \frac{\hbar^2 \nabla^2 {\psi} (\mr, t) }{2m}
 + g {\psi}^\dagger (\mr, t)
{\psi}^2 (\mr, t)
\end{equation}
by the replacement
$$
{\psi} (\mr, t) \rightarrow \Phi (\mr, t)
= \langle {\psi} (\mr, t) \rangle_{\rm coh.}
$$
where the expectation value is taken with a coherent state of
the particle creation and annihilation operator
of the lowest single particle state.
This approximation is also known as the tree approximation in the language of
the loop expansion\cite{RJ74} and ignores quantum fluctuations around the mean
field $\Phi (\mr, t)$.
It is the purpose of the present work to investigate the effect of quantum corrections
to the mean field by the variational method which we have developed for
describing
dynamics of quantum fields.\cite{TVM99,TVM00}

When the effective interaction of the form (\ref{1.1}) is used to calculate the effect of the
quantum fluctuations, a special care is needed to avoid (ultraviolet) divergences
which arises due to the singular nature of the potential which causes coupling
of all momentum modes.\cite{LHY57,TTH97,BN97}
It has been shown by Lee, Huang and Yang\cite{LHY57} that such divergences
can be eliminated systematically by the proper use of the pseudo-potential
defined by\cite{BW52}
\begin{equation}
V_{\rm ps.} (\mr) = g \delta  (\mr) \frac{\partial}  {\partial r} r
= g \delta  (\mr) + g \delta  (\mr) r \frac{\partial}  {\partial r} \ .
\end{equation}
The difference,
\begin{equation}
\Delta V (\mr) = V_{\rm ps.} (\mr) - V_{\rm eff.} (\mr)
= g \delta  (\mr) r \frac{\partial}  {\partial r} \ ,
\end{equation}
which has only vanishing contribution in the first order, generates divergent
terms in higher
order and it is precisely these divergent terms which cancel the divergences
which are contained in the higher order quantum corrections of
$V_{\rm eff.} (\mr)$.\cite{LHY57}

Physically this procedure is understood as the elimination of the double
counting in the usual perturbation expansion in terms of the bare two particle
interaction $V_0 (\mr)$:
since the pair scattering (in vacuum) is already fully accounted for by the
scattering length $a$, one should not include,
when performing many body calculation, contributions from repeated
use of the effective interaction (\ref{1.1})  in the two particle
scattering channel.\footnote{Although the procedure to eliminate the ultraviolet 
divergence found in the literature\cite{FW71,LP78} is very suggestive, 
this interpretation of {\it the subtraction of the double counting terms} was never 
explicitly mentioned elsewhere to the best of our knowledge.}
For example, the contributions from the diagram in Fig. 1 (a)
should not be included
in the many body calculation, because it is a part of the diagrams Fig. 1 (b)
which is already included in the lowest order diagrams Fig. 1 (c)
in terms of $V_{\rm eff.}$. 
The problem of ultraviolet divergence can of course be avoided if one starts from a bare two-body 
interaction of finite range; this approach would however involve much more elaborate 
calculations even for a static system.\cite{TBLD00}

Elimination of the ultraviolet divergences due to the singular potential
looks technically similar to the renormalization of the effective
local couplings in quantum field theories to eliminate ultraviolet
divergences.\cite{LP78}
This analogy was exploited further by Toyoda\cite{TT81} and
more recently by Braaten and Nieto.\cite{BN97}
However, here we choose to remove all divergences by simply avoiding double countings
of pair scatterings, by inspection, based on the observation that
the effective interaction (\ref{1.1}) should be used only
in the first Born ``approximation" to describe two particle scattering.

\begin{figure}[ht]
  \vskip -2cm
  \epsfxsize=12cm  
  \centerline{\epsfbox{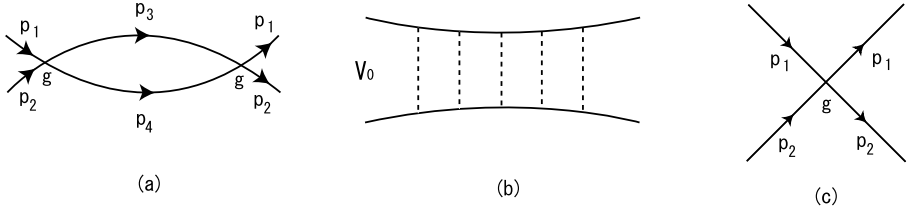}}
  \caption{The one-loop diagram depicted in (a) should not be included in
  the calculation since it is a part of (b) which is already included
  in the first order (tree) diagram (c).}
   \label{fig1}
\end{figure}

\begin{figure}[ht]
  \vskip -3cm
  \epsfxsize=8cm  
  \centerline{\epsfbox{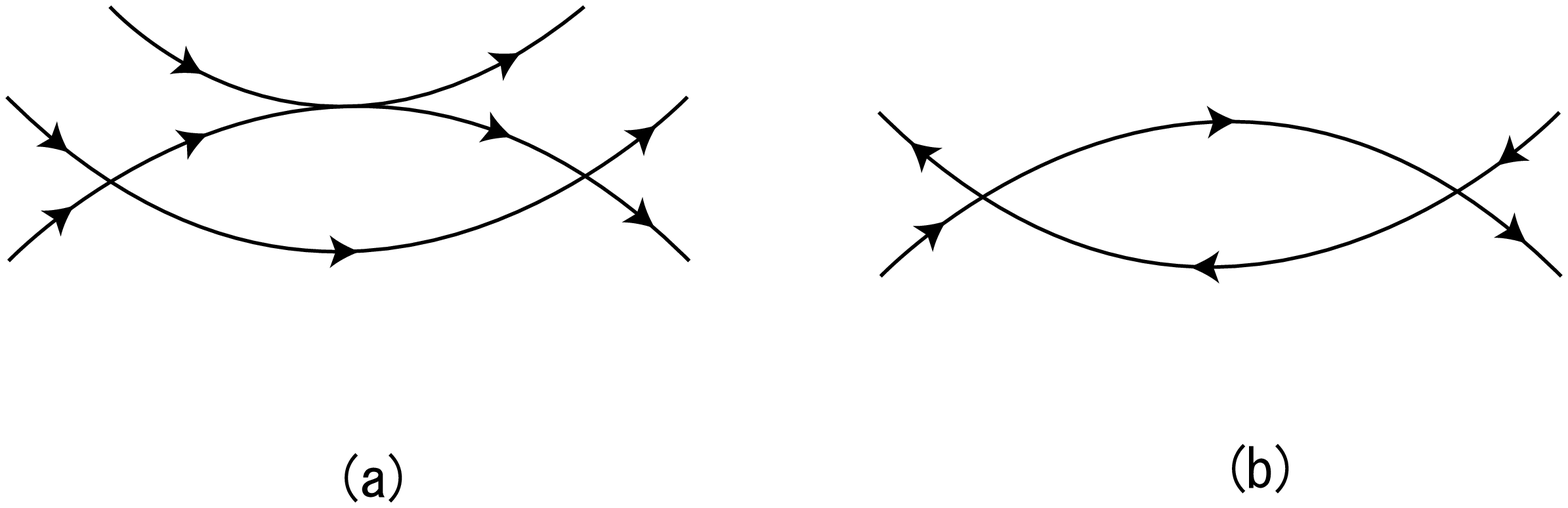}}
  \caption{Some diagrams which cannot be reduced to the pair
    interaction like Fig. 1 (b) and therefore should be included in
    the calculation. }
   \label{fig2}
\end{figure}

This does not mean, however, that we should discard the quantum fluctuations
entirely: the many-body interaction such as Fig. 2 should be still
retained in the calculation.  For example, the well-known Bogoliubov
theory\cite{NNB47} contains
the effect of quantum fluctuations as zero-point oscillations of
quasi-particle modes;
the contribution to the energy density of this fluctuation would
diverge if one uses
the potential of the form (\ref{1.1}).
This divergence can be eliminated by the above procedure
leaving behind non-analytic $a^{5/2}$ correction which originates
from the sum of genuine many-body interaction contributions.\cite{LHY57}

In this paper we present the basic formulation of the variational method
and derive the equations of motion which generalizes the Gross-Pitaevskii 
mean field equation (\ref{GPeq}) to include the effect of quantum fluctuations
in terms of the reduced density matrix.
The resultant equations are similar to those known as the Hartree-Fock-Bogoliubov 
equations for fermions in nuclear many-boby theory\cite{RS80} and are 
identical, although slightly disguised, to the ones obtained for interacting 
bose-systems by Blaizot and Ripka\cite{BR86}, and more recently 
by Griffin\cite{AG96}. 
The general static solutions of these equations are given in terms of the mode 
functions.
We then perform an explicit calculation for a uniform condensate and show how 
the divergent terms appear by the use of the effective potential (\ref{1.1}).
By the inspection of the origin of these divergences in perturbation series
we will demonstrate that all these divergences can be eliminated by the systematic
removal of the terms which correspond to the double counting of pair scattering.
By this method we obtain a gap equation free of divergence, (\ref{3-37}), a new 
result obtained in this paper: 
To the best of our knowledge, this equation has never been written explicitly 
in the literature.
In the forthcoming paper, we shall apply the present method to compute the
density response function of the system.

The rest of the paper is organized as follows.
In the next section, we construct the Gaussian variational wave functional
for condensates in a general trapping external potential. 
In section 3, we derive the equations of motion which consist of a generalized Gross-Pitaevskii 
equation for the condensate and the Liouville equation for the reduced density matrix. 
Our generalized Gross-Pitaevskii equations consist of two coupled
non-linear equations: one is
the equation of motion for $\Phi (\mr, t)$ similar to the
Gross-Pitaevskii equation (\ref{GPeq}),
but it also contains the mean field terms associated
with the fluctuations, and the other is the equation of motion for
the fluctuations, written in the form of the Liouville equation
for the reduced density matrix consisting of fluctuations,
these equations are often called the time-dependent Hartree-Bogoliubov
equation.
In section 4, we show that these equations can be easily extended to finite
temperatures using
the Gaussian density matrix which is determined self-consistently.
In section 5, we express the static solutions of these equations in terms 
of the mode functions.
In section 6 we examine the static solution in uniform system in detail and 
compare the results to the well-known Bogoliubov's  mean field theory.
We illustrate how the divergence can be eliminated by the removal of
the double counting terms in the diagrams of perturbative series. 
A brief summary is given in section 7.

\def\mib#1{\mbox{\boldmath $#1$}}
\newcommand{\bra}[1]{\langle {#1} |}    
\newcommand{\ket}[1]{| {#1} \rangle}     

\section{Gaussian wave functional}
In this and following sections we shall apply the variational method developed
earlier\cite{KV89,TVM99}
in the functional Schr\"{o}dinger picture to derive an extended form of
the Gross-Pitaevskii equation
which contain the effect of quantum fluctuations in a minimal fashion.

We consider a non-relativistic gas of bose particles of mass $m$ interacting via
contact interaction in the form of (\ref{1.1}).   The dynamics of system is described
by the following second quantized Hamiltonian:
\begin{eqnarray}\label{2-1}
& & {\hat H }=\int d^3{\mib x}
\biggl\{-\frac{1}{2m}\psi^{\dagger}({\mib x})\nabla^2
\psi({\mib x})
+\psi^{\dagger}({\mib x})  V_{\rm ext}({\mib x}) \psi({\mib x})
\nonumber\\
& &\qquad\qquad\qquad
+\frac{g}{2}\psi^{\dagger}({\mib x})\psi^{\dagger}({\mib x})
\psi({\mib x})\psi({\mib x})\biggl\} \ .
\end{eqnarray}
Here, $V_{\rm ext}$ represents an external confining potential
provided by the magnetic trap and the mutual interaction of particle
is given by the effective interaction of the form (\ref{1.1}) .
The fields $\psi({\mib x})$ are quantized by the 
commutation relation :
$[\ \psi({\mib x}) \ , \ \psi^{\dagger}({\mib y}) \ ]
=\delta^3({\mib x}-{\mib y})$ and
the other combinations are equal to 0.
Hereafter, we take $\hbar=1$.

In order to apply the Gaussian wave functional method, which has been developed
in our previous work\cite{TVM99}, Schr\"odinger picture,
we need to rewrite the Hamiltonian in terms of canonical coordinates and their
associate momenta.
There is some earlier attempt to define such coordinates\cite{HT96}, but we choose 
here the representation which diagonalize the bilinear part of the Hamiltonian.
\begin{equation}
{\hat H}_0  \equiv \int d^3{\mib x}
\biggl\{-\frac{\hbar^2}{2m}\psi^{\dagger}({\mib x})\nabla^2 \psi({\mib x})
+\psi^{\dagger}({\mib x})V_{\rm ext}({\mib x})\psi({\mib x}) \biggl\}
= \sum_\alpha  \omega_\alpha a_\alpha^\dagger  a_\alpha
\end{equation}
The field operator is decomposed as 
\begin{equation}
\psi (\mx) = \sum_\alpha \varphi_\alpha (\mx) a_\alpha
\end{equation}
where $\varphi_\alpha (\mx)$ is the normalized eigen wave function
of free bosons in external potential.
We then introduce the following Hermitian operators :
\begin{eqnarray}\label{2-3}
& & Q_\alpha
\equiv \frac{1}{\sqrt{2\omega_\alpha}} (a_\alpha + a^{\dagger}_\alpha ) \ ,
\nonumber\\
& & P_\alpha
\equiv - i \sqrt{\frac{\omega_\alpha}{2}}
(a_\alpha - a^{\dagger}_\alpha) \ .
\end{eqnarray}
With $\int d \mx \varphi_\alpha^* (\mx) \varphi_\beta (\mx) = \delta_{\alpha, \beta}$
these operators satisfy the canonical commutation relations:
$$[\ Q_\alpha \ , \ P_\beta \ ]
= i \delta_{\alpha, \beta} \ ,$$
an analogue to that of the position and momentum operator
in quantum mechanics,
and convert the free Hamiltonian into a sum of uncoupled harmonic oscillators:
\begin{equation}
{\hat H}_0 = \sum_\alpha \frac{1}{2} \left( P_\alpha^2 + \omega_\alpha^2 Q_\alpha^2 \right) \ .
\end{equation}
In the $Q$-representation, corresponding to the
coordinate representation in quantum mechanics, the field momentum
operator is expressed by the functional differential: $P_\alpha =
-i \frac{\delta}{\delta Q_\alpha}$.
The vacuum state of the Hamiltonian ${\hat H}_0$ is expressed as
 a product of Gaussian:
\begin{eqnarray}
\Psi_0 [\phi] = \la \phi | 0 \ra
& = & {\cal N}_0 \exp \left[ - \frac{1}{2} \sum_\alpha  \omega_\alpha Q_\alpha^2 \right]
\nonumber \\
&=& {\cal N}_0 \exp \left[ - \frac{1}{2} \int d^3 \mx d^3 \my
\phi^\dagger ( \mx )  W_0 ( \mx, \my ) \phi (\my) \right]
\end{eqnarray}
where in the last line we have used 
the complex $\phi$-field defined by\footnote{Do not confuse $\phi (\mx )$
with the original bose field $\psi (\mx )$: it obeys the commutation relation,
$ [ \phi (\mx) , \pi^\dagger (\my) ] = i \delta (\mx - \my) $
where the $\pi$-field is defined by
$ \pi ( \mx) = \sum_\alpha \varphi_\alpha (\mx) P_\alpha$. }
$$ \phi ( \mx) = \sum_\alpha \varphi_\alpha (\mx) Q_\alpha$$
and the kernel of the integral is defined by
\begin{equation}
W_0 ( \mx, \my ) = \sum_\alpha \varphi_\alpha (\mx) \omega_\alpha \varphi^*_\alpha (\my) \ .
\end{equation}
Indeed, this state will be annihilated when we operate the particle annihilation operator $a_\alpha$ from the left
for any $\alpha$.
In the uniform system ($V_{\rm ext.} = V_0$) :
$\varphi_\alpha (\mx) \rightarrow \frac{1}{\sqrt{V}} e^{i {\bf k} \cdot \mx}$,
$\omega_\alpha \rightarrow k^2/2m + V_0 $,
and $W_0 ( \mx, \my ) \rightarrow \delta (\mx - \my )
(-\nabla_\mx^2 / 2m + V_0)$.

Our time-dependent variational wave functional in the functional Schr\"odinger
picture is written as a straightforward generalization of the time-dependent
Gaussian wave function in quantum mechanics:\cite{TVM99}
\begin{eqnarray}\label{2-4}
\Psi[\phi] &=&
{\cal N}\exp \left[ 
i\langle{\bar \pi}|\phi-{\bar \phi}\rangle 
-\bra{\phi-{\bar \phi}}\frac{1}{4G}+i\Sigma
\ket{\phi-{\bar \phi}} \right] 
\end{eqnarray}
where we have used abbreviate notations for integrals such as
\begin{eqnarray*}
\la {\bar \pi}|\phi - {\bar \phi} \ra
& = &\sum_\alpha {\bar P}_\alpha (t) \left( Q_\alpha - {\bar Q}_\alpha (t)
\right)
= \int d^3 \mx
{\bar \pi}^* (\mx,t ) \left( \phi (\mx ) - {\bar \phi} (\mx, t ) \right) \ , \\
\bra{\phi} K \ket{\phi} 
& = & \sum_{\alpha,\beta} Q_\alpha K_{\alpha, \beta} (t) Q_\beta
= \int d^3 \mx d^3 \my \phi^* (\mx) K (\mx , \my; t)  \phi (\my)  \
\end{eqnarray*}
for $K = G^{-1}, \Sigma$.
Here two (complex) functions,
$${\bar \phi} (\mx, t) = \sum_\alpha \varphi_\alpha (\mx) {\bar Q}_\alpha (t)
\ , \qquad {\bar \pi} (\mx, t) = \sum_\alpha \varphi_\alpha (\mx) {\bar P}_\alpha (t)
\ , $$
and two kernels,
\begin{eqnarray*}
G^{-1} (\mx , \my; t) & = &
\sum \varphi_\alpha (\mx) G^{-1}_{\alpha, \beta} (t)
\varphi^*_\beta (\my )  \\
\Sigma (\mx , \my; t) & = &
\sum \varphi_\alpha (\mx) \Sigma_{\alpha, \beta} (t)
\varphi^*_\beta (\my )
\end{eqnarray*}
are to be considered as time-dependent variational parameters of
the wave functional.

The expectation value of a composite operator ${\cal O} (\phi,\pi) $
is given by the functional integral
$$
\langle {\cal O} \rangle = \int [ d \phi ] \Psi^*[\phi] {\cal O}
\left(\phi, -i \frac{\delta}{\delta \phi} \right) \Psi [\phi]
$$
which becomes a functional of the variational parameters
${\bar \phi}({\mib x}, t)$, ${\bar \pi}({\mib x}, t)$, $G({\mib x}, {\mib y}, t)$
and $\Sigma({\mib x},{\mib y},t)$.
For example, the expectation values of the field $\phi (\mx)$ and its canonical momentum
$\pi (\mx)$ are given by
\begin{eqnarray}
\la \phi (\mx) \ra & = & {\bar \phi}({\mib x},t)
\\
\la \pi (\mx) \ra & = & {\bar \pi}({\mib x},t)
\end{eqnarray}
respectively, and
\begin{eqnarray}\label{meanfield}
\langle \psi({\mib x}) \rangle
& = & \sum_\alpha \sqrt{\frac{\omega_\alpha}{2}} \varphi_\alpha (\mx)
\left( {\bar Q}_\alpha (t)
+ \frac{i}{\omega_\alpha} {\bar P}_\alpha (t) \right)
\equiv 
\Phi ({\mib x},t) \ , \\
\langle \psi^{\dagger} ({\mib x}) \rangle
& = & \sum_\alpha \sqrt{\frac{\omega_\alpha}{2}} \varphi_\alpha (\mx)
\left( {\bar Q}_\alpha (t) - \frac{i}{\omega_\alpha} {\bar P}_\alpha (t)
\right)
\equiv 
\Phi^*({\mib x},t) \ .
\end{eqnarray}
The nonvanishing expectation value of the particle creation
and annihilation operators implies that our wave functional is not
an eigen state of the particle number.  Our Gaussian state corresponds
to a squeezed state, or a generalized coherent state, in
quantum optics.

In order to describe field fluctuations it is convenient to introduce
\begin{equation}
{\hat \psi} (\mx, t) \equiv \psi (\mx) - \Phi (\mx, t ) \ .
\end{equation}
For example, we obtain
\begin{equation}\label{2-6}
\langle \psi^{\dagger}({\mib x})\psi({\mib y})\rangle
= \Phi^* (\mx,t) \Phi (\my, t)
+\langle {\hat \psi}^{\dagger}({\mib x},t){\hat \psi}({\mib y},t)\rangle
\end{equation}
and so on, where
$\langle {\hat \psi}^{\dagger}({\mib x},t){\hat \psi}({\mib y},t)\rangle$
is the two-point correlation function at time $t$ and represents
the quantum fluctuation.
The fluctuation is determined by the width parameter of the Gaussian
wave functional.
The explicit forms may be obtained from the following relations:\cite{TVM99}
\begin{eqnarray}
\la \hQ_\alpha \hQ_\beta \ra & = & G_{\alpha, \beta} (t) \\
\la \hP_\alpha \hP_\beta \ra
& = & \frac{1}{4} G^{-1}_{\alpha, \beta} (t) + 4 \left( \Sigma G \Sigma \right)_{\alpha, \beta} (t) \\
\la \hQ_\alpha \hP_\beta \ra
& = & -2 \left( G \Sigma \right) _{\alpha, \beta} (t) + \frac{1}{2} i \delta_{\alpha, \beta} \\
\la \hP _\alpha \hQ_\beta \ra
& = & -2 \left( \Sigma G \right) _{\alpha, \beta} (t) - \frac{1}{2} i \delta_{\alpha, \beta}
\end{eqnarray}
where $\hQ_\alpha  = Q_\alpha - {\bar Q}_\alpha$ and $\hP_\alpha  = P_\alpha - {\bar P}_\alpha$.
For example,
\begin{eqnarray*}
\la \hphi (\mx) \hphi^\dagger (\my) \ra & = &
\sum_{\alpha, \beta} \varphi_\alpha (\mx) \la \hQ_\alpha \hQ_\beta \ra \varphi^*_\beta (\my)
\nonumber \\
& = & \sum_{\alpha, \beta} \varphi_\alpha (\mx)  G_{\alpha, \beta} (t) \varphi^*_\beta (\my)
= G ( \mx,\my ; t )
\end{eqnarray*}
and similar relations can be obtained for
$\la \hphi^\dagger (\mx) \hpi (\my) \ra $ etc.
However, we do not need such explicit forms in the following discussion.

With the Gaussian form of the variational wave functional, expectation
value of all higher products of the field operators are expressed in terms
of the center (the mean field) and the width
(the two-point correlation function)
of the Gaussian.
This feature, characteristic of the Gaussian Ansatz of the fluctuations,
enables one to truncate otherwise infinite
set of coupled equations of the motion for the expectation values of
the products of the quantum fields
into a set of coupled equations of motion for the mean field
$\Phi (\mx, t)$ and the two-point functions of fluctuations
$G(\mx , \my ; t)$.

\section{Equations of motion}

The equations of motion 
can be derived from the time-dependent variational principle as applied
to the action:
\begin{equation}\label{action}
S = \int dt \la \Psi | i \frac{\partial }{\partial t} - {\hat H}
+ \mu {\hat N} | \Psi \ra
\ ,
\end{equation}
where $\mu$ is the chemical potential of the bosons introduced to enforce that the
total number of the bosons $N = \la \Psi | {\hat N} | \Psi \ra$, defined with
$ {\hat N} = \int d \mx \psi^\dagger ( \mx ) \psi ( \mx )$, is conserved.
Minimization of (\ref{action}) with respect to the variational parameters
leads to the equations of motion of ${\bar \phi} (\mx, t)$ and
${\bar \pi} (\mx, t) $,
$G^{-1} (\mx , \my; t)$ and $\Sigma (\mx , \my; t)$\cite{TVM99}
which can be rewritten for the mean field and the fluctuations.

Alternatively, one can derive the equations of motion written in terms of the
mean field and fluctuation directly from:
$i{\dot{\Phi}}
={\delta \langle {\hat H}' \rangle}/{\delta ({\Phi}^*)}$ and
$i{\dot{\Phi}}^*
=-{\delta \langle {\hat H}' \rangle}/{\delta {\Phi}}$.
Here, $\langle {\hat H}' \rangle$ is the expectation value of
the effective Hamiltonian ${\hat H}' = {\hat H} - \mu {\hat N}$.
Then one finds:
\begin{eqnarray}\label{gGPeq}
i{\dot{\Phi}}({\mib x},t)\nonumber & = &
-\frac{1}{2m}\nabla^2{\Phi}({\mib x},t)
+ \left( V_{\rm ext}({\mib x}) - \mu \right) {\Phi}({\mib x},t)
+g|{\Phi}({\mib x},t)|^2{\Phi}({\mib x},t) \nonumber\\
& & \qquad +2g\langle {\hat \psi}^\dagger({\mib x})
{\hat \psi}({\mib x})\rangle {\Phi}({\mib x},t)
+g\langle {\hat \psi}({\mib x})
{\hat \psi}({\mib x})\rangle {\Phi}^*({\mib x},t) \ ,
\end{eqnarray}
which is a generalized form of the Gross-Pitaevskii equation (\ref{GPeq})
including the effect of fluctuations.

Our next task is to derive equations of motion for the two-point functions:
$\langle{\hat \psi}^\dagger (\mx) {\hat \psi} (\my) \rangle$ and
$\langle{\hat \psi} (\mx){\hat \psi} (\my)\rangle$.
This can be done from the equations of motion of the kernel functions
$G(\mx,\my,t)$ and $\Sigma (\mx,\my,t)$.
For this purpose we introduce the $2 \times 2$ reduced density matrix
defined as
\begin{eqnarray}\label{rdm}
{\cal M}({\mib x},{\mib y},t)+\frac{1}{2}\delta^3({\mib x}-{\mib y})
=
\pmatrix{
\langle {\hat \psi}({\mib x}){\hat \psi}^{\dagger}({\mib y})\rangle
 &
-\langle {\hat \psi}({\mib x}){\hat \psi}({\mib y})\rangle \cr
\langle {\hat \psi}^{\dagger}({\mib x})
{\hat \psi}^{\dagger}({\mib y})\rangle &
-\langle {\hat \psi}^{\dagger}({\mib x}){\hat \psi}({\mib y})\rangle
} \ .
\end{eqnarray}
Then one can show that the equation of motion for
the reduced density matrix ${\cal M}$
is expressed in the form of the Liouville equation:
\begin{equation}\label{LVeq}
i{\dot {\cal M}}= [ \ {\cal H} \ , \ {\cal M}\ ] \ ,
\end{equation}
where the Hamiltonian density ${\cal H}$ is defined by
\begin{equation}\label{H}
{\cal H}({\mib x},{\mib y},t) \equiv \delta^3({\mib x}-{\mib y})
\pmatrix{ W({\mib x}) & \Delta({\mib x}) \cr
          -\Delta^*({\mib x}) & -W^*({\mib x}) }  \
\end{equation}
with
\begin{eqnarray}\label{W}
W({\mib x}) & \equiv &
-\frac{1}{2m}\nabla_{\mib x}^2+V_{\rm ext}({\mib x}) - \mu
+2g|{\Phi}({\mib x},t)|^2  
+2g\langle {\hat \psi}^{\dagger}({\mib x}){\hat \psi}({\mib x})\rangle
\nonumber
\\
& = & W^* ({\mib x})
\\ \ \
\label{Delta}
\Delta({\mib x})  & \equiv &
g({\Phi}({\mib x},t)^2
+\langle {\hat \psi}({\mib x}){\hat \psi}({\mib x})\rangle) \ . 
\end{eqnarray}
As we shall show, this equation reproduces the quasiparticle excitations
in the Bogoliubov theory when the term induced by the fluctuations
are discarded.

Equations (\ref{gGPeq}) and (\ref{LVeq}) with (\ref{meanfield}),
(\ref{rdm}), (\ref{H}),
(\ref{W}), and (\ref{Delta}), form a closed set of equations
which describe non-linear
time evolution of the Bose-Einstein condensate and its fluctuations
described by the Gaussian state.
They are an extension of the Gross-Pitaevskii equation to include
quantum fluctuations.

\section{Inclusion of thermal fluctuations}

Foregoing derivation based on the pure state (\ref{2-4})
can be extended to include thermal fluctuations by introducing
statistical average with a Gaussian density matrix
for mixed states.\cite{EJP88}
If the system is near equilibrium, it is equivalent to introduce
the density operator in the form:
\begin{equation}\label{stat}
D (t) = \frac{1}{Z}e^{-\beta {\overline H} (t) } \ , \qquad \hbox{\rm with} \quad
Z={\rm Tr} e^{-\beta{\overline H} (t) } \ ,
\end{equation}
where $\beta=1/k_{\rm B}T$ is the inverse temperature and ${\overline H} (t)$
is the (time-dependent) mean field Hamiltonian given by
\begin{eqnarray}\label{HBh}
{\overline H} (t) = \int d^3{\mib x}\biggl\{& &
-\frac{1}{2m}{\psi}^{\dagger}({\mib x})\nabla^2{\psi}({\mib x})
+(V_{\rm ext}({\mib x})-\mu)
{\psi}^{\dagger}({\mib x}){\psi}({\mib x})
\nonumber\\
& &+\frac{g}{2}\Bigl[
\langle \psi^{\dagger}({\mib x})\psi^{\dagger}({\mib x})\rangle
{\psi}({\mib x}){\psi}({\mib x})
+\langle \psi({\mib x})\psi({\mib x})\rangle
{\psi}^{\dagger}({\mib x}){\psi}^{\dagger}({\mib x}) \nonumber\\
& &\qquad\qquad\qquad\qquad
+4\langle \psi^{\dagger}({\mib x})\psi({\mib x})\rangle
{\psi}^{\dagger}({\mib x}){\psi}({\mib x})
\Bigl]\biggl\} \ .
\end{eqnarray}
Here,
$\langle \psi^{\dagger}({\mib x})\psi({\mib x})\rangle
=\Phi^*(\mx,t)\Phi(\mx,t)
+\langle {\hat \psi}^{\dagger}({\mib x}){\hat \psi}({\mib x})\rangle$,
$\langle \psi^{\dagger}({\mib x})\psi^{\dagger}({\mib x})\rangle
=\Phi^*(\mx,t)\Phi^*(\mx,t)
+\langle {\hat \psi}^{\dagger}({\mib x})
{\hat \psi}^{\dagger}({\mib x})\rangle$ and
$\langle \psi({\mib x})\psi({\mib x})\rangle
=\Phi(\mx,t)\Phi(\mx,t)
+\langle {\hat \psi}({\mib x})
{\hat \psi}({\mib x})\rangle$.
The mean field and the fluctuations are now defined by
\begin{eqnarray}
\Phi ( \mx, t ) & = &
\la \psi ( \mx ) \ra = {\rm Tr} \left( D (t) \psi ( \mx ) \right) \\
\la \hphi ( \mx ) \hphi (\my) \ra
& = & {\rm Tr} \left( D (t) \hphi ( \mx ) \hphi (\my)  \right)  \qquad \hbox{\rm etc.}
\label{Dpsipsi}
\end{eqnarray}

One can derive the equation of motion from the Liouville equation of the non-equilibrium
density matrix $D$,
\begin{equation}
i \frac{\partial D}{\partial t}  = \left[ {\overline H} , D \right] \ .
\end{equation}
For example, the equation of motion of the mean field can be obtained from
\begin{eqnarray}
i \frac{\partial \Phi}{\partial t}  & = & i {\rm Tr} \left( \dot{D} (t) \psi ( \mx ) \right)
={\rm Tr} \left( \left[ {\overline H} , D \right]  \psi ( \mx )  \right)
={\rm Tr} \left( D \left[ \psi ( \mx ),  {\overline H}  \right] \right)
\ .
\end{eqnarray}
Upon insertion of the (\ref{HBh}) to compute the commutator $\left[ {\overline H} , \psi ( \mx ) \right] $
we reproduce the generalized Gross-Pitaevskii equation in the form as (\ref{gGPeq}) which
we have obtained for the pure state evolution.
Similarly, by replacing $\psi (\mx)$ by $\hpsi (\mx) \hpsi (\my)$, etc. one can derive the equation of motion
for the reduced density matrix ${\cal M}$ defined by (\ref{rdm}) with (\ref{Dpsipsi}) which becomes
precisely the same as the equation (\ref{LVeq}).

We can therefore reinterpret that the equations, (\ref{gGPeq}) and (\ref{LVeq}), also describe the non-equilibrium
time evolution
\footnote{These equations describe the coherent, adiabatic evolution of the system, while kinetic 
aspects (e.g. the growth of the condensate\cite{GZ97,S99,ZNG99}) are beyond the scope of the present method. }
of the perturbation introduced in the system which has been in equilibrium at temperature $T$
and the chemical potential $\mu$.

\section{Static solutions}

We first consider the case in which the mean field $\Phi$ and the fluctuation ${\cal M}$ do not
depend on time, that is, the system is in the thermal equilibrium.
In this case, the equation of motion for ${\cal M}$ becomes
\begin{equation}\label{2-10}
[ \ {\cal H} \ , \ {\cal M}\ ]=0 \ .
\end{equation}
This implies that the Hamiltonian density $ {\cal H} $ and the reduced density matrix ${\cal M}$
can be simultaneously diagonalized.  This is achieved explicitly by introducing the
mode functions.

The mode functions, $u({\mib x},t)$ and $v({\mib x},t)$, for the Hamiltonian matrix
${\cal H}$ are defined by:
\begin{equation}\label{2-12}
i\partial_t \pmatrix{ u({\mib x},t) \cr v({\mib x},t)}
=
\pmatrix{ W({\mib x}) & \Delta({\mib x}) \cr
          -\Delta^*({\mib x}) & -W^*({\mib x}) }
          \pmatrix{ u({\mib x},t) \cr v({\mib x},t)} \ .
\end{equation}
For the stationary state, we write
\begin{equation}\label{2-13}
u({\mib x},t)=e^{-iEt}u({\mib x}) \ , \qquad
v({\mib x},t)=e^{-iEt}v({\mib x}) \ .
\end{equation}
The eigenvalue equation is then obtained as
\begin{equation}\label{2-14}
\pmatrix{ W({\mib x}) & \Delta({\mib x}) \cr
          -\Delta^*({\mib x}) & -W^*({\mib x}) }
          \pmatrix{ u_\alpha({\mib x}) \cr v_\alpha({\mib x})}
=E_\alpha \pmatrix{ u_\alpha({\mib x}) \cr v_\alpha({\mib x})} \ ,
\end{equation}
where $\alpha$ represents quantum numbers.
By taking a complex conjugate of the above eigenvalue equation,
it can be shown that the following equation is also satisfied :
\begin{equation}\label{2-15}
\pmatrix{ W({\mib x}) & \Delta({\mib x}) \cr
          -\Delta^*({\mib x}) & -W^*({\mib x}) }
          \pmatrix{ v_\alpha^*({\mib x}) \cr u_\alpha^*({\mib x})}
=-E_\alpha \pmatrix{ v_\alpha^*({\mib x}) \cr u_\alpha^*({\mib x})} \ .
\end{equation}
Namely, there exist negative energy solutions.
The eigenvector $(u, v)$ is also the eigenvector for the reduced
density matrix ${\cal M}$ because ${\cal H}$ and ${\cal M}$ commute :
\begin{eqnarray}\label{2-16}
\int d^3{\mib y} \left(
{\cal M}({\mib x},{\mib y})+\frac{1}{2}\delta^3({\mib x}-{\mib y})
\right)
\pmatrix{ u_\alpha({\mib y}) \cr v_\alpha({\mib y})}
=(f_\alpha({\mib x})+1/2)
\pmatrix{ u_\alpha({\mib x}) \cr v_\alpha({\mib x})} \ . \qquad
\end{eqnarray}
where, by construction (\ref{rdm}),
the eigenvalues of ${\cal M}$ are related to the quasi-particle distribution
$n_\alpha$ by\cite{TVM99,TVM00}
\begin{equation}\label{2-17}
f_\alpha=n_\alpha + \frac{1}{2} \ , \qquad
n_{\alpha}=\frac{1}{e^{\beta E_{\alpha}}-1} \ .
\end{equation}
Taking the complex conjugate of the eigenvalue equation
for ${\cal M}+1/2$, we also obtain
\begin{eqnarray}\label{2-18}
\int d^3{\mib y} \left(
{\cal M}({\mib x},{\mib y})+\frac{1}{2}\delta^3({\mib x}-{\mib y})
\right)
\pmatrix{ v_\alpha^*({\mib y}) \cr u_\alpha^*({\mib y})}
=-(f_\alpha({\mib x})-1/2)
\pmatrix{ v_\alpha^*({\mib x}) \cr u_\alpha^*({\mib x})} \ . \qquad
\end{eqnarray}
Here the ortho-normalization relation is given by
\begin{equation}\label{2-19}
\int d^3{\mib x}\left(
u_\alpha^*({\mib x})u_\beta({\mib x})
-v_{\alpha}^*({\mib x})v_\beta({\mib x})\right)
=\pm \delta_{\alpha\beta} \ ,
\end{equation}
where $\alpha>0\ (<0)$ and $\beta>0\ (<0)$ corresponds to
$+\ (-)$ on the right-hand side and
it is implied that $\alpha>0\ (<0)$ represents the positive (negative) energy
solution.
The reduced density matrix ${\cal M}$ can be
expressed in terms of the mode functions $u$ and $v$ as
\begin{eqnarray}\label{2-20}
{\cal M}({\mib x},{\mib y})+\frac{1}{2}\delta^3({\mib x}-{\mib y})
&=&\sum_{\alpha>0}\biggl\{
\pmatrix{u_\alpha({\mib x}) \cr v_\alpha({\mib x})}
(f_\alpha({\mib x})+1/2)
\pmatrix{u_\alpha^*({\mib y}),  & \!\! -v_\alpha^*({\mib y})} \nonumber\\
& &\ \
-\pmatrix{v_\alpha^*({\mib x}) \cr u_\alpha^*({\mib x})}
(f_\alpha({\mib x})-1/2)
\pmatrix{-v_\alpha({\mib y}),  & \!\! u_\alpha({\mib y})}
\biggl\} \ . \qquad\quad
\end{eqnarray}
Here, summation is taken only over the positive energy eigenstates.
This result implies, from 
the definition (\ref{rdm}),
the matrix elements of the reduced density matrix
are expressed in terms of the mode functions and
the bose distribution function $n_{\alpha}$ :
\begin{eqnarray}\label{2-21}
& &\langle {\hat \psi}^{\dagger}({\mib x}){\hat \psi}({\mib y})\rangle
=\sum_{\alpha>0}[
(n_\alpha({\mib x})+1)v_\alpha({\mib x})v_\alpha^*({\mib y}) 
+n_\alpha({\mib x})u_\alpha^*({\mib x})u_\alpha({\mib y})]
\nonumber\\
& &\langle {\hat \psi}({\mib x}){\hat \psi}({\mib y})\rangle
=\sum_{\alpha>0}[
(n_\alpha({\mib x})+1)u_\alpha({\mib x})v_\alpha^*({\mib y}) 
+n_\alpha({\mib x})v_\alpha^*({\mib x})u_\alpha({\mib y})]
\nonumber\\
& &\langle {\hat \psi}^{\dagger}({\mib x}){\hat \psi}^{\dagger}
({\mib y})\rangle
=\sum_{\alpha>0}[
(n_\alpha({\mib x})+1)v_\alpha({\mib x})u_\alpha^*({\mib y}) 
+n_\alpha({\mib x})u_\alpha^*({\mib x})v_\alpha({\mib y})]
\nonumber\\
& &\langle {\hat \psi}({\mib x}){\hat \psi}^{\dagger}({\mib y})\rangle
=\sum_{\alpha>0}[
(n_\alpha({\mib x})+1)u_\alpha({\mib x})u_\alpha^*({\mib y}) 
+n_\alpha({\mib x})v_\alpha^*({\mib x})v_\alpha({\mib y})] \ .
\end{eqnarray}
The equations (\ref{gGPeq}), (\ref{2-14}) and (\ref{2-21})
with (\ref{2-17}) form a 
basic set of equations
for the static Bose-Einstein condensates
which need to be solved self-consistently.

\section{Uniform condensate}
In this section, we consider a simple uniform system, neglecting external confining potential,
$V_{\rm ext}=0$, in order to illustrate how our method works.
In this case, the condensate is spatially uniform:
${\Phi}({\mib x})={\Phi}_0$ and all mode functions are reduced to 
simple plane waves.  The indices $\alpha$
are regarded as momenta ${\mib k}$ and the mode
functions $u$ and $v$ may be written as
\begin{equation}\label{3-1}
u_{\alpha} \longrightarrow u_{\bf k}({\bf x})
= \frac{1}{\sqrt{V}} e^{i{\bf k}\cdot\mx} u_{\bf k} \ , \qquad
v_{\alpha} \longrightarrow v_{\bf k}({\bf x})
= \frac{1}{\sqrt{V}}e^{i{\bf k}\cdot\mx} v_{\bf k} \ ,
\end{equation}
where $V$ is the volume of the system.
Further, we take the continuum limit ($V \longrightarrow \infty$) with the
prescription
$\sum_{\bf k} \longrightarrow {V}/{(2\pi)^3}\cdot\int d^3{\bf k}$
and 
$\delta_{\alpha\beta} \longrightarrow \delta({\bf k}-{\bf k}')$.

The generalized Gross-Pitaevskii equation now takes the form,
\begin{equation}\label{3-2}
(-\mu+g|\Phi_0|^2+2g\langle{\hat \psi}^\dagger{\hat \psi}
\rangle){\Phi_0}+g\langle{\hat \psi}{\hat \psi}\rangle
{\Phi}^*_0=0 \ ,
\end{equation}
where
\begin{eqnarray}
& &
\langle{\hat \psi}^{\dagger}{\hat \psi}\rangle
= 
 \int \frac{d^3 \bf k}{(2\pi)^3} [ (n_{\bf k}+1)|v_{\bf k}|^2
+n_{\bf k}|u_{\bf k}|^2 ] \ ,
\label{3-3}
\\
& &\langle{\hat \psi}{\hat \psi}\rangle
= 
 \int \frac{d^3 \bf k}{(2\pi)^3}
(2n_{\bf k}+1) u_{\bf k}v_{\bf k}^*
 \  \label{psipsi}
\end{eqnarray}
and the mode functions and their eigenvalues
are easily obtained from (\ref{2-14}) as 
\begin{eqnarray}\label{3-4}
& &
u_{\bf k}=
\sqrt{\frac{W_{\bf k}+E_{\bf k}}{2E_{\bf k}}}
\ , \quad
v_{\bf k}=-
\sqrt{\frac{W_{\bf k}-E_{\bf k}}{2E_{\bf k}}}
\frac{\Delta^*}{|\Delta|} \  
\end{eqnarray}\label{spectrum}
with 
\begin{equation}\label{Ek}
E_{\bf k} = \sqrt{W_{\bf k}^2-|\Delta|^2} \ ,
\end{equation}
where
\begin{eqnarray}\label{3-5}
W_{\bf k} & = & \frac{{\mib k}^2}{2m}-\mu+2g( |{\Phi}_0|^2
+\langle{\hat \psi}^{\dagger}{\hat \psi}\rangle) \ ,
\\
\Delta & = & g({\Phi_0}^2+\langle{\hat \psi}{\hat \psi}\rangle) \ .
\label{delta}
\end{eqnarray}

Inserting (\ref{psipsi}) into (\ref{delta}) and then using (\ref{3-4}),
we obtain the ``gap equation" for the anomalous density:
\begin{eqnarray}\label{gapeq}
\frac{\Delta}{g}  = \langle \psi \psi \rangle & =  &
\Phi_0^2 + \langle \hpsi \hpsi \rangle
\nonumber \\
&= & \Phi_0^2  - \Delta \int \frac{d^3 {\bf k}}{(2\pi)^3}
\frac{n_{\bf k}+1/2}{E_{\bf k}} \ .
\end{eqnarray}
On the other hand, the particle number density is given by
\begin{eqnarray}
n   =  \langle \psi^\dagger \psi \rangle & = & 
|\Phi_0|^2 + \langle{\hat \psi}^{\dagger} {\hat \psi} \rangle
\nonumber \\
\label{3-18}
& = & |\Phi_0|^2 + \int \frac{d^3{\mib k}}{(2\pi)^3}
\left( n_{\bf k} \frac{W_{\bf k}}{E_{\bf k}} 
+ \frac{W_{\bf k}-E_{\bf k}}{2E_{\bf k}} \right) \ ,
\end{eqnarray}
where we have used (\ref{3-4}) again.
Equations, (\ref{3-2}) and (\ref{gapeq}), together with the constraint
(\ref{3-18}) and
with the quasi-particle distribution function (\ref{2-17}),
determine $\Delta$, $\Phi_0$ and $\mu$, self-consistently for a given value of
particle density $n$ and temperature $T$.

We can show that the well-known Bogoliubov spectrum of the quasi-particle can
be reproduced from (\ref{Ek}) by removing all fluctuations.
For simplicity, we consider the ground state,
$n_{\bf k}=0$ for all ${\mib k}\neq 0$
and assume that
${\Phi}_0$ is real. 
Then, by setting
$\langle {\hat \psi}^\dagger {\hat \psi} \rangle
= \langle {\hat \psi}{\hat \psi} \rangle
= \langle {\hat \psi}^\dagger {\hat \psi}^\dagger \rangle = 0$, we find
\begin{equation}\label{3-7}
\mu = | \Delta | = g | \Phi_0 |^2 \ , 
\end{equation}
and 
\begin{eqnarray}\label{3-8}
E_{\bf k}&=& \sqrt{W_{\bf k}^2-|\Delta|^2}
= \sqrt{\frac{{\mib k}^2}{2m}
\left(\frac{{\mib k}^2}{2m}+2g |\Phi_0|^2\right)} \ ,
\end{eqnarray}
where we have used
\begin{equation}
W_{\bf k} = \frac{{\mib k}^2}{2m}- \mu +2g | \Phi_0|^2
= \frac{{\mib k}^2}{2m} + g |\Phi_0|^2 \ ,
\end{equation}
which precisely coincides with
the Bogoliubov dispersion relation for quasi-particle excitations.
For small $|{\mib k}|$, with $g > 0$,  the phonon dispersion relation
$\omega=ck$ is obtained, where the sound velocity $c$ is
\begin{equation}\label{3-9}
c=\sqrt{\frac{g | \Phi_0 |^2}{m}} \ .
\end{equation}
This is the well-known result of the Bogoliubov theory.

The appearance of the phonon dispersion relation is a consequence of
spontaneous breakdown of the continuous $U(1)$ symmetry of the Hamiltonian
associated with the
phase change of the bose field:
$\psi ( \mx ) \to e^{i \delta} \psi ( \mx )  $.  The
phonon can be regarded as a Nambu-Goldstone mode of the broken
$U(1)$ (or $O(2)$) symmetry.

Our formula (\ref{Ek}) for the spectrum of the mode functions
with the Gaussian fluctuation, however,
does not reproduce phonon spectrum, but instead exhibits a gap at
$ {\bf k} \rightarrow 0$, violating
Goldstone's theorem.
This shortcoming is rather inherent in the Gaussian wave functional approach,
\footnote{According to the classification of Hohenberg and Martin\cite{HM65},
our approximation corresponds to the conserving (or $\Phi$-derivable\cite{GB62}) 
approximation, as opposed to the gapless approximation\cite{gapless}.  
The origin of the violation of 
the Goldstone theorem is traced back to the "Fock" term which may be suppressed
by keeping the leading term in the $1/N$ expansion\cite{CJP74} for a system with O(N) 
symmetry.\cite{BS03}}
but this defect can be remedied by computing the excitation spectrum.\cite{TVM00}

We now turn our discussion on the removal of divergences. 
The expression  (\ref{psipsi}) for the fluctuations in the reduced density matrix
actually contain a divergent integral, and so does the gap equation
(\ref{gapeq}).
To see this explicitly, we observe that at large
$| {\bf k} |$, 
\begin{eqnarray}
E_{\bf k} \simeq W_{\bf k} -  \frac{| \Delta |^2}{2 W_{\bf k}}
\end{eqnarray}
so that
\begin{eqnarray}
u_{\bf k}   \simeq 1 + \frac{|\Delta|^2}{4 W_{\bf k}^2} ,  \qquad
v_{\bf k}  \simeq  - \frac{\Delta^*}{2 W_{\bf k}} \ .
\end{eqnarray}
Inserting this expression in (\ref{psipsi}), we find that the off-diagonal
component of the density matrix becomes
\begin{equation}
\langle{\hat \psi}{\hat \psi}\rangle
=  \int \frac{d^3 \bf k}{(2\pi)^3}
(2n_{\bf k}+1) u_{\bf k}v_{\bf k}^*
\simeq  - \int \frac{d^3 \bf k}{(2\pi)^3}
(2n_{\bf k}+1) \frac{\Delta}{2 W_{\bf k}}  \ ,
\end{equation}
which indeed contains a linearly divergent integral,
\begin{equation}
\langle{\hat \psi}{\hat \psi}\rangle_{\rm div.} =
-\Delta \int \frac{d^3 \bf k}{(2\pi)^3} \frac{m}{ {\bf k}^2} \ .
\end{equation}

\begin{figure}[t]
  \vskip -3cm
  \epsfxsize=6cm  
  \centerline{\epsfbox{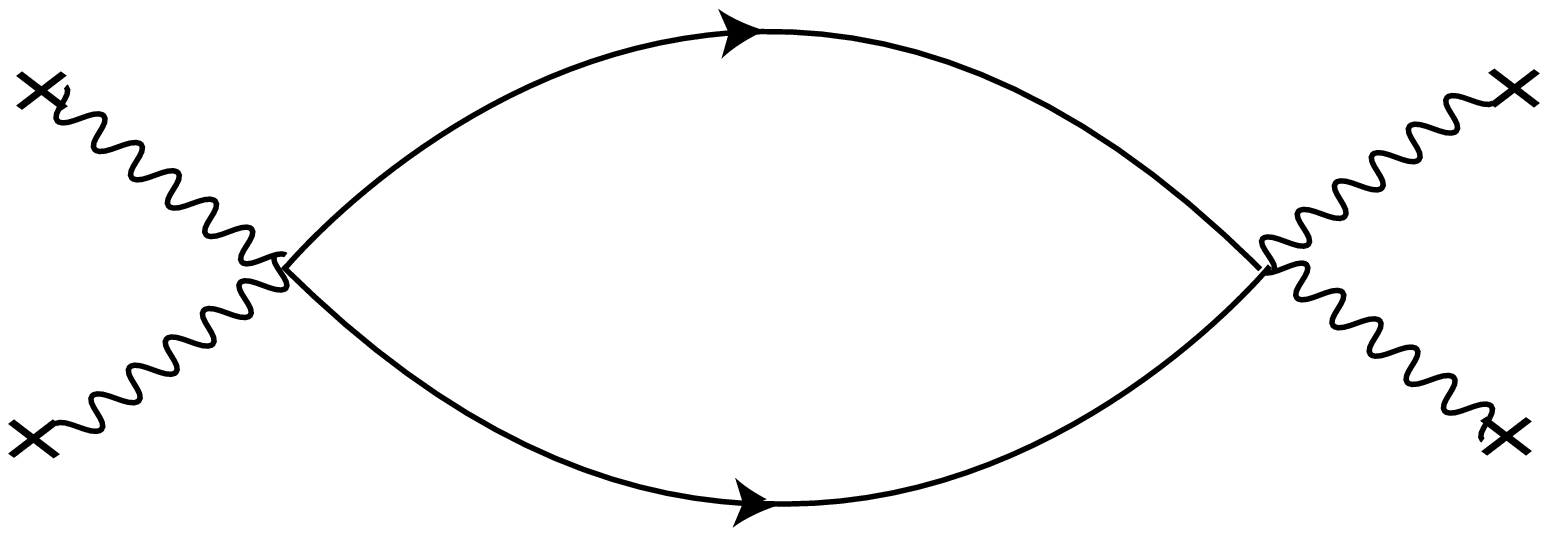}}
  \caption{The divergent contribution of the off-diagonal component
  of the density matrix.}
   \label{fig3}
\end{figure}

This divergent integral is associated with the diagram shown in Fig. 3.
It originates from a repeated use of the effective
potential as we have mentioned in the introduction.
This divergence should be removed in order to avoid the double
counting of the perturbation series in terms of the bare potential.
Therefore we define the renormalized density matrix by subtracting
the divergent term:
\begin{equation}
\langle{\hat \psi}{\hat \psi}\rangle_{\rm ren.} =
\langle{\hat \psi}{\hat \psi}\rangle -
\langle{\hat \psi}{\hat \psi}\rangle_{\rm div.}
\end{equation}
This procedure is equivalent to replace the gap equation by:
\begin{equation}\label{3-37}
\frac{\Delta}{g} =  \Phi_0^2  -  \Delta \int \frac{d^3 {\bf k}}{(2\pi)^3}
\left( \frac{n_{\bf k}+1/2}{E_{\bf k}} \ - \frac{m}{{\bf k}^2 } \right) \ .
\end{equation}
With this replacement and all other relations untouched,
all divergences associated with double counting of the pair scatterings
are removed from the previous equations. 
The "renormalized gap equation" (\ref{3-37}) is one of our new findings in this 
paper.

\section{Summary}
In this paper we have formulated the variational method to describe the dynamics of the 
Bose-Einstein condentate interacting via short-range effective interaction.  
With the Gaussian variational wave functional we have derived the equations of motion 
which consist of a generalized Gross-Pitaevskii equation for the condensate and a 
Liouville-von Neumann equation for the density matrix of the quantum and thermal 
fluctuations.  The static solutions of these equations are examined in terms of the
mode functions for the fluctuations.  In the case of uniform condensate, these mode
functions are reduced to simple plane waves, and we are able to perform the calculation
explicitly.  We have shown that the spectrum of the mode functions reduces to the 
Bogoliubov phonon spectrum when we neglect the effect of quantum fluctuations;
it does not contain however phonon mode in the presence of quantum fluctuation. 
In these calcuations we have shown that the divergent integrals which arise due to 
the use of singular effective interaction can be eliminated systematically 
by removing the repeated pair scattering terms to avoid the double counting 
of the diagrams in terms of the bare interaction.  
In the forthcoming paper,  we will show how to apply the present method to compute 
the linear response of the system to external perturbation and to extract the phonon dispersion 
relation of the collective excitations in the sytem.

\acknowledgements
We thank Prof. Gordon Baym for calling our attention to the ref. \cite{LP78} concerning the physical 
interpretation of the removal of divergences.  One of us (T. M.) is grateful to Jean-Paul Blaizot for 
very helpful discussions from which we were benefitted to revising the original version of the manuscript.
This work is supported in part by the Grants-in-Aid for the Scientific Research No. 13440067
and No. 13740159 from the Ministry of Education, Culture, Sports, Science and Technology in Japan.

\appendix

\end{document}